\documentclass[aip,jcp,amsmath,amssymb,reprint,graphicx]{revtex4-1}

\usepackage{graphicx}
\usepackage{dcolumn}
\usepackage{bm}
\usepackage{physics}
\usepackage[utf8]{inputenc}
\usepackage[T1]{fontenc}
\usepackage{mathptmx}

\begin{document}

\title{Real-time density-matrix coupled-cluster approach for closed and open systems at finite temperature}

\author{Philip Shushkov}
\email[]{shushkov@caltech.edu}
\author{Thomas F. Miller III}
\affiliation{Division of Chemistry and Chemical Engineering, California Institute of Technology, Pasadena, California 91125, USA}

\date{\today}

\begin{abstract}
We extend the coupled-cluster method to correlated quantum dynamics of both closed and open systems at finite temperatures using the thermo-field formalism. The approach expresses the time-dependent density matrix in an exponential ansatz and describes time-evolution along the Keldysh path contour. A distinct advantage of the approach is exact trace-preservation as a function of time, ensuring conservation of probability and particle number. Furthermore, the method avoids the computation of correlated bra-states, simplifying the computational implementation. We develop the method in a thermal quasi-particle representation, which allows seamless connection to the projection method and diagrammatic techniques of the traditional coupled-cluster formalism. For comparison, we also apply the thermo-field framework to renormalization-group methods to obtain reference results for closed and open systems at finite temperatures. We test the singles and doubles approximation to the density-matrix coupled-cluster method on the correlated electronic dynamics of the single-impurity Anderson model, demonstrating that the new method successfully captures the correlated dynamics of both closed systems at finite temperature and driven-dissipative open systems. This encouraging performance motivates future applications to non-equilibrium quantum many-body dynamics in realistic systems.
\end{abstract}


\maketitle

\section{Introduction}
The simulation of non-equilibrium quantum many-body dynamics is an outstanding challenge in chemical physics. The theoretical description of non-equilibrium phenomena is essential for the understanding of many processes, including chemical dynamics of molecules,\cite{krausz2009attosecond,domcke2012role} defects in extended systems,\cite{doherty2013nitrogen,falk2015optical,freysoldt2014first} quantum transport in molecular junctions,\cite{thoss2018perspective,cui2017perspective} and chemical reactions at surfaces and interfaces,\cite{zhang2017surface,jiang2019dynamics,maurer2016thermal} as well as the description of novel driven-dissipative phenomena in open systems, such as discrete time crystals.\cite{choi2017observation,else2019discrete} These open questions motivate the extension of well-established wavefunction approaches of quantum chemistry to correlated quantum dynamics of closed and open systems at finite temperature.

Among the methods of quantum chemistry, the coupled-cluster approach has become the golden standard for incorporation of correlation effects in molecular systems.\cite{bartlett2007coupled,crawford2000introduction} The approach relies on an exponential ansatz for the wavefunction of the system, which is parameterized by connected cluster amplitudes. The singles and doubles approximation to the cluster operator together with a perturbative triples correction has set the accuracy standard for ground state properties of molecules, and excited state properties have been targeted via linear response theory.\cite{bartlett2012coupled,nooijen1997new,krylov2008equation} Furthermore, time-dependent coupled-cluster wavefunction methods have been recently applied to the non-equilibrium electronic dynamics of molecules,\cite{huber2011explicitly,sato2018communication,kvaal2012ab} and beyond the single-reference framework, the coupled-cluster approach has been extended to more challenging, multi-reference systems.\cite{lyakh2011multireference}

Thermal effects have been recently included in the coupled-cluster formalism\cite{white2018time,hummel2018finite} using the method of thermal cumulants.\cite{sanyal1992thermal,sanyal1993systematic,mandal2003finite} Alternatively, thermo-field theory has been combined with the method of configuration interaction to simulate finite-temperature molecular properties.\cite{harsha2019thermofield} The thermo-field method\cite{takahashi1996thermo,chu1994unified} was developed to incorporate thermal mixed states in quantum field theory and the approach was later extended to open systems.\cite{arimitsu1985general,arimitsu1987non,arimitsu1988dissipative,arimitsu1987general} Besides configuration interaction treatment, the method has allowed direct application of coupled-cluster theory to non-equilibrium stationary states,\cite{dzhioev2011super,dzhioev2012nonequilibrium,dzhioev2011second,dzhioev2014nonequilibrium,dzhioev2014superoperator} and it has been also combined with renormalization-group methods\cite{de2015thermofield,schwarz2018nonequilibrium,borrelli2019density,borrelli2016quantum,borrelli2017simulation} to simulate the real-time finite-temperature dynamics of closed systems, including impurity-bath models and coupled electron-nuclear dynamics. 

Our goal is to extend the coupled-cluster approach to the real-time correlated quantum dynamics of closed and open systems at finite temperatures, bridging the gap between recent finite-temperature and real-time coupled-cluster extensions. We provide a unified framework to describe both equilibrium and non-equilibrium stationary states and non-equilibrium quantum dynamics. We use the thermo-field theory to represent the mixed quantum states of closed and open systems in a wavefunction form, and we express this wavefunction in an exponential ansatz. We test the singles and doubles approximation to the new density-matrix coupled-cluster method on the correlated electronic dynamics of the single-impurity Anderson model\cite{newns1969self,anderson1961localized,wilson1975renormalization} and compare the results with the renormalization-group method. 

The paper is organized as follows. We summarize thermo-field theory and present the real-time finite-temperature density-matrix coupled-cluster method for closed and open systems in Sec.~\ref{theor}. We present the results for the quantum dynamics of the single-impurity Anderson model at a range of interaction strengths and temperatures in Sec.~\ref{results}, including also dynamics with driving and dissipation. We conclude in Sec.~\ref{conclusion} with remarks on the future directions for method development. 

\section{Theory}\label{theor}
 We start this section by reviewing the thermo-field formalism and the thermal Bogoliubov transformation, which we use to define temperature-dependent quasi-particles. We then present the main result of this paper, the extension of the real-time coupled-cluster method to closed and open systems at finite temperatures using the thermal quasi-particle representation. We focus next on the test system for our quantum dynamics approximations, the single-impurity Anderson model, and the extension of renormalization group approaches to closed and open systems, which we compare with the coupled-cluster approximation.

\subsection{Thermo-field theory}
Thermo-field theory\cite{takahashi1996thermo,arimitsu1985general,arimitsu1987general,schmutz1978real} expresses the mixed states of quantum systems in the form of pure state wavefunctions. Let us define a basis in the Hilbert space $\mathcal{H}$ of a system by the ket-states $\ket{\boldsymbol{n}}$, generated via the action of the creation operators $a^{\dagger}_i$ on the vacuum ket $\ket{0}$, $\ket{\boldsymbol{n}} = \prod_i a^{\dagger}_i \ket{0}$, where $\boldsymbol{n}$ is the set of occupation numbers and the index $i$ enumerates the states of the single-particle basis. The creation operators $a^{\dagger}_i$ and their Hermitian conjugate annihilation operators $a_i$ obey the commutation relations
\begin{equation}\label{comm}
    \left[a_i,a^{\dagger}_j\right]_\sigma = \delta_{ij}, \left[a^{\dagger}_i,a^{\dagger}_j\right]_\sigma = \left[a_i,a_j\right]_\sigma = 0,
\end{equation}
where the commutator is $\left[A,B\right]_\sigma = AB - \sigma BA$ with $\sigma=1$ for bosons and $\sigma=-1$ for fermions. Using the basis $\ket{\boldsymbol{n}}$ of $\mathcal{H}$, the density matrix $\rho$ is given by
\begin{equation}\label{rho}
\rho = \sum_{n,m} \rho_{nm} \ket{\boldsymbol{n}} \bra{\boldsymbol{m}}
\end{equation}
with the ket-bra couples $\ket{\boldsymbol{n}} \bra{\boldsymbol{m}}$ forming an orthonormal basis. To write Eq.~(\ref{rho}) in the form of a wavefunction, thermo-field theory uses a second replica $\tilde{\mathcal{H}}$ of the original Hilbert space $\mathcal{H}$ to express the ket-bra couples $\ket{\boldsymbol{n}} \bra{\boldsymbol{m}}$ as ket-basis states $\ket{\boldsymbol{n},\tilde{\boldsymbol{m}}}$, effectively doubling the number of degrees of freedom of the system. We define tilde creation $\tilde{a}^{\dagger}_i$ and annihilation $\tilde{a}_i$ operators in the tilde Hilbert space $\tilde{\mathcal{H}}$ via the relations
\begin{equation}\label{tdef}
\begin{split}
    \tilde{a}_i \ket{\boldsymbol{n}} \bra{\boldsymbol{m}} &= \sigma^{\mu+1} \ket{\boldsymbol{n}} \bra{\boldsymbol{m}} a^{\dagger}_i, \\
    \tilde{a}^{\dagger}_i \ket{\boldsymbol{n}} \bra{\boldsymbol{m}} &= \sigma^\mu \ket{\boldsymbol{n}} \bra{\boldsymbol{m}} a_i
\end{split}
\end{equation}
with $\mu=\sum_i (n_i-m_i)$, the difference of ket- and bra-state occupation numbers. From Eqs.~(\ref{comm}) and ~(\ref{tdef}), the non-tilde and tilde operators satisfy the commutation relations
\begin{equation}\label{tcomm}
\begin{split}
    \left[\tilde{a}_i,\tilde{a}^{\dagger}_j\right]_\sigma = \delta_{ij}, \left[\tilde{a}^{\dagger}_i,\tilde{a}^{\dagger}_j\right]_\sigma & = \left[\tilde{a}_i,\tilde{a}_j\right]_\sigma = 0, \\
    \left[\tilde{a}^{\dagger}_i,a^{\dagger}_j\right]_\sigma & = \left[\tilde{a}_i,a_j\right]_\sigma = 0,
\end{split}
\end{equation}
which show that the non-tilde and tilde operators represent independent fermionic or bosonic degrees of freedom. Given that the tilde Hilbert space is Hermitian conjugate to the original Hilbert space, thermo-field theory relates operators acting on the two subspaces via the anti-linear tilde conjugation operation with the properties
\begin{equation}\label{tildedef}
\begin{split}
     \left(AB\right)^{\tilde{}} &= \tilde{A} \tilde{B}, \\
     \left(c_a A + c_b B\right)^{\tilde{}} &= c^{*}_a \tilde{A} + c^{*}_b \tilde{B},
\end{split}
\end{equation}
with $c_a$ and $c_b$ complex numbers and $*$ denoting complex conjugation. The double tilde conjugation rules $\tilde{\tilde{a}}_i = \sigma a_i$ and $\tilde{\tilde{a}}^{\dagger}_i = \sigma a^{\dagger}_i$ supplement Eq.~(\ref{tildedef}).

In the doubled Hilbert space $\mathcal{H} \cross \tilde{\mathcal{H}}$ of non-tilde (physical) and tilde (hidden) degrees of freedom, the density matrix, Eq.~(\ref{rho}), takes the form of a pure state wavefunction
\begin{equation}\label{rhoket}
    \rho = \sum_{n,m} \rho_{nm} \ket{\boldsymbol{n},\tilde{\boldsymbol{m}}}.
\end{equation}
The thermal density matrix $\ket{\rho_0}$ of a non-interacting system, for instance, can be written in the exponential form
\begin{equation}\label{urhoth}
    \ket{\rho_0} = \frac{1}{Q_0} \sum_n e^{-\beta  \sum_i n_i \epsilon_i} \ket{\boldsymbol{n},\tilde{\boldsymbol{n}}} = \frac{1}{Q_0} \text{exp} \left(\sum_i e^{-\beta \epsilon_i} a^{\dagger}_i \tilde{a}^{\dagger}_i \right) \ket{0},
\end{equation}
where $\beta$ is the reciprocal temperature, $\beta=(k_BT)^{-1}$ with $T$ the temperature, $Q_0$ is the partition function, $\epsilon_i$ are the single-particle energy levels, and $\ket{0} = \ket{0,\tilde{0}}$ is the vacuum ket-state in the doubled Hilbert space. Similarly, the resolution of identity, the unit ket-state $\ket{1}$, also has an exponential form
\begin{equation}\label{unit}
    \ket{1} = \sum_n \ket{\boldsymbol{n},\tilde{\boldsymbol{n}}} = \text{exp} \left( \sum_i a^{\dagger}_i \tilde{a}^{\dagger}_i \right) \ket{0}.
\end{equation}

The average value of an operator $O$, given by the trace with the density matrix $\rho$, becomes an expectation value with the ket-state $\ket{\rho}$, Eq.~(\ref{rhoket}), in thermo-field theory. Specifically,
\begin{equation}\label{expval}
    \text{tr} \left[\rho O \right] = \bra{1} O \ket{\rho},
\end{equation}
where the density matrix is normalized $\bra{1}\ket{\rho}=1$. The bra-state and ket-state in Eq.~(\ref{expval}) are different due to the choice of the path contour in the complex time plane as shown in Sec.~\ref{cc}. This choice is most suitable for extending the coupled-cluster approach to finite temperatures because it circumvents the need to compute a correlated bra-state. We provide further details in Sec.~\ref{cc} devoted to the coupled-cluster formalism.

The Liouville-von Neumann equation for the time-evolving density matrix, $i\Dot{\rho} = \left[H,\rho\right]$, becomes Schr\"odinger equation for the pure ket-state $\ket{\rho}$
\begin{equation}\label{se}
    i\ket{\Dot{\rho}_t} = \hat{H} \ket{\rho_t},
\end{equation}
where the super-Hamiltonian operator $\hat{H}$ is the difference of the original Hamiltonian $H$ and its tilde conjugate
\begin{equation}\label{sham}
    \hat{H} = H - \tilde{H}.
\end{equation}
The super-Hamiltonian in Eq.~(\ref{sham}) does not mix the non-tilde and tilde degrees of freedom for a closed system, resulting in a unitary evolution. For open, dissipative systems, the super-Hamiltonian entangles the non-tilde and tilde degrees of freedom by terms that resemble pairing of opposite spin electrons in the theory of superconductivity. We will give an example of such a super-Hamiltonian in Sec.~\ref{siam}.

Formal integration of Eq.~(\ref{se}) gives the time-dependent density matrix $\ket{\rho_t}$ as
\begin{equation}\label{rhot}
    \ket{\rho_t} = e^{-i \hat{H} t} \ket{\rho_{t_0}}
\end{equation}
with $\ket{\rho_{t_0}}$ the initial density matrix. In quantum-quench simulations, we initiate the quantum dynamics from an uncorrelated state and take into account the effects of interactions by the subsequent time evolution. In Sec.~\ref{results}, we present results of such simulations where most of the system, the bath, is initially at thermal equilibrium with a density matrix given by Eq.~(\ref{urhoth}). Non-equilibrium states relax to stationary states, such as equilibrium states, with time-independent density matrices $\ket{\rho_s}$, which are annihilated by the super-Hamiltonian $\hat{H} \ket{\rho_s}=0$.

\subsection{Thermal quasi-particles}\label{tbt}
The creation $a^{\dagger}_i$ and $\tilde{a}^{\dagger}_i$ and annihilation $a_i$ and $\tilde{a}_i$ operators have a complementary action on the unit bra-state
\begin{equation}\label{brarel}
\begin{split}
    \bra{1} a^{\dagger}_i &= \bra{1} \tilde{a}_i, \\
    \bra{1} \tilde{a}^{\dagger}_i &= \sigma \bra{1} a_i.
\end{split}
\end{equation}
Because the super-Hamiltonian, Eq.~(\ref{sham}), conserves the number of particles and contains equal numbers of non-tilde and tilde operators, Eq.~(\ref{brarel}) gives
\begin{equation}
    \bra{1} \hat{H} = 0.
\end{equation}
This bra-state condition ensures the conservation of probability and the normalization of the time-dependent density matrix, Eq.~(\ref{rhot}), at all times, and it allows us to express the expectation value, Eq.~(\ref{expval}), in the Heisenberg picture
\begin{equation}
    \bra{1} O \ket{\rho_t} = \bra{1} O_t \ket{\rho_{t_0}}
\end{equation}
with $O_t = e^{i \hat{H} t} O e^{-i \hat{H} t}$ -- the Heisenberg-evolved operator $O$.

The action of the $a_i$ and $\tilde{a}_i$ operators on the non-interacting thermal density matrix is similarly complementary
\begin{equation}\label{ketrel}
\begin{split}
    a_i \ket{\rho_0} &= e^{-\beta \epsilon_i} \tilde{a}^{\dagger}_i \ket{\rho_0}, \\
    \tilde{a}_i \ket{\rho_0} &= \sigma e^{-\beta \epsilon_i} a^{\dagger}_i \ket{\rho_0}.
\end{split}
\end{equation}
Because $a_i$ annihilates an electron, the thermal ket-state condition, Eq.~(\ref{ketrel}), shows that $\tilde{a}^{\dagger}_i$ creates a hole, and vice versa because $a^{\dagger}_i$ creates an electron, $\tilde{a}_i$ annihilates a hole.

Eqs.~(\ref{brarel}) and ~(\ref{ketrel}) imply that we can transform the creation $a^{\dagger}_i$ and $\tilde{a}^{\dagger}_i$ and annihilation $a_i$ and $\tilde{a}_i$ operators to a new set of creation $b^{\dagger}_i$ and $\tilde{b}^{\dagger}_i$ and annihilation $b_i$ and $\tilde{b}_i$ operators, which respectively annihilate the unit bra-state and non-interacting thermal ket-state
\begin{equation} \label{newvac}
    \begin{split}
        \bra{1} b^{\dagger}_i &= \bra{1} \tilde{b}^{\dagger}_i = 0, \\
        b_i \ket{\rho_0} &= \tilde{b}_i \ket{\rho_0} = 0.
    \end{split}
\end{equation}
The transformation that satisfies Eq.~(\ref{newvac}) and preserves the commutation relations, Eqs.~(\ref{comm}) and ~(\ref{tcomm}), is given by 
\begin{equation} \label{btrans}
    \begin{split}
    b^{\dagger}_i = a^{\dagger}_i - \tilde{a}_i, &\;\; \tilde{b}^{\dagger}_i = \tilde{a}^{\dagger}_i - \sigma a_i, \\
    b_i = u_i a_i - v_i \tilde{a}^{\dagger}_i, &\;\; \tilde{b}_i = u_i \tilde{a}_i - \sigma v_i a^{\dagger}_i,
    \end{split}
\end{equation}
where $v_i = \left( e^{\beta \epsilon_i} - \sigma \right)^{-1}$ is the Bose-Einstein or the Fermi-Dirac distribution and $u_i = 1 + \sigma v_i$. Eq.~(\ref{btrans}) preserves the commutation relations among the new set of creation and annihilation operators but does not preserve the Hermitian conjugation relation between creation and annihilation operators. Such canonical, non-unitary transformations are known in the theory of superconductivity as Bogoliubov transformations.\cite{blaizot1986quantum} The thermal Bogoliubov transformation, Eq.~(\ref{btrans}), defines new thermal quasi-particles $b^{\dagger}_i$ and $b_i$ and thermal quasi-holes $\tilde{b}^{\dagger}_i$ and $\tilde{b}_i$ with respect to the unit bra-state and non-interacting thermal ket-state, Eq.~(\ref{newvac}), which we refer to as the new left $\bra{0}$ and right $\ket{0}$ vacuum states. We use the thermal quasi-particle representation to introduce correlated dynamics via coupled-cluster theory at finite temperatures in closed and open systems in the following section.

\subsection{Density-matrix coupled-cluster approach}\label{cc}
In our coupled-cluster approach, we express the time-dependent density matrix of the system in the exponential form
\begin{equation}\label{dmcc}
    \ket{\rho_t} = e^{T_t} \ket{\rho_{t_0}},
\end{equation}
where $\ket{\rho_{t_0}}=\ket{\rho_{0}}$ is the non-interacting thermal density matrix in Eq.~(\ref{urhoth}), and $T_t$ is the time-dependent cluster operator, which we expand in the number of excited thermal quasi-particle-hole pairs $n$ as
\begin{equation}\label{tser}
    T_t = \sum^{N}_{n=1} T_n,
\end{equation}
where $N$ is the highest excitation level. We express the cluster operators $T_n$ via the quasi-particle $b^{\dagger}_i$ and quasi-hole $\tilde{b}^{\dagger}_i$ creation operators from Sec.~\ref{tbt} as
\begin{equation}\label{tbop}
    T_n = \frac{1}{n!} \sum_{ij ... , kl ... } t^{ik...}_{jl...} b^{\dagger}_i b^{\dagger}_k ... \tilde{b}^{\dagger}_l \tilde{b}^{\dagger}_j,
\end{equation}
where $t^{ik...}_{jl...}$ are the time-dependent cluster amplitudes, and the indeces $i$, $j$, $k$, and $l$ run over the entire single-particle basis. We test the performance of the density-matrix coupled-cluster (DMCC) method keeping terms with up to two quasi-particle-hole pairs in Eq.~(\ref{tser}). The explicit form of the cluster operator for this singles and doubles approximation is
\begin{equation}\label{dmccsd}
    T_t = \sum_{ij} t^i_j b^{\dagger}_i \tilde{b}^{\dagger}_j + \frac{1}{4} \sum_{ijkl} t^{ik}_{jl} b^{\dagger}_i b^{\dagger}_k \tilde{b}^{\dagger}_l \tilde{b}^{\dagger}_j.
\end{equation}

Eq.~(\ref{dmcc}) together with Eqs.~(\ref{tser}) and ~(\ref{tbop}) parameterize a class of trace-preserving transformations that include the subset of dynamical completely positive transformations of open systems and the narrower subset of unitary transformations of closed systems. The trace-preservation directly follows from the definition of the $b^{\dagger}$ and $\tilde{b}^{\dagger}_i$ creation operators, Eq.~(\ref{newvac}), $\bra{1}\ket{\rho_t} = \bra{1} e^{T_t} \ket{\rho_0} = \bra{1}\ket{\rho_0}$. The time evolution governed by the Schr\"odinger equation, Eq.~(\ref{se}), imposes further constraints on the density matrix in addition to trace-preservation, such as the Hermiticity and positivity of the density matrix. These constraints translate to relations among the cluster amplitudes. The Hermiticity of the density matrix, for instance, gives invariance upon tilde conjugation in thermo-field theory, $\ket{\tilde{\rho_t}} = \ket{\rho_t}$, resulting in a tilde invariant cluster operator $\tilde{T_t} = T_t$. For a singles-only approximation, this leads to the Hermiticity of the singles amplitude matrix $t^i_j = t^{j*}_i$ and similar relations hold for the higher rank cluster amplitudes. The dynamical constraints determined by the Schr\"odinger equation need not be incorporated in the parameterization of the cluster amplitudes; rather, their satisfaction may be used as diagnostics for the solutions to the Schr\"odinger equation.

The trace-preservation property of the coupled-cluster density matrix allows us to express quantum-statistical averages of time-dependent operators in a simple way.\cite{matsumoto1984equivalence} Thermal averages of Heisenberg-evolved operators $O(t)$ are given in general by the ratio
\begin{equation}\label{green}
    G = \frac{\text{tr} \left[e^{-\beta H} O(t_N) .. O(t_1)\right]}{\text{tr} \left[e^{-\beta H}\right]} = \frac{\langle T_c U_c O_N .. O_1 \rangle}{\langle T_c U_c \rangle}.
\end{equation}
In diagrammatic terms, the denominator in Eq.~(\ref{green}) cancels the disconnected vacuum diagrams in the numerator, such that only connected diagrams contribute to the thermal average. In Eq.~(\ref{green}), we have also introduced a more succinct expression for the average that involves a time-ordered propagation with $T_c$ the chronological operator and $U_c$ the full quantum propagator along a contour in the complex time plane, which initiates at time $t_i$ before time $t_1$, evolves along the real time axis to $t_f$ past the time $t_N$, comes back along the real time axis to $t_i$, and propagates downwards along the imaginary time axis to time $t_i-i\beta$. We can take respectively the initial $t_i$ and final $t_f$ times to the infinite past, $t_i \rightarrow -\infty$, and infinite future, $t_f \rightarrow \infty$, where the interaction is turned off, resulting in cancellation of the correlated contribution to the imaginary axis propagation. Such a contour in the complex time plane underlies non-equilibrium Keldysh diagrammatic theory\cite{keldysh1965diagram,altland2010condensed} and the density-matrix coupled-cluster theory presented here. 

Using the definition of expectation value Eq.~(\ref{expval}) and the super-Hamiltonian Eq.~(\ref{sham}), we cast Eq.~(\ref{green}) in thermo-field theory as
\begin{equation}\label{greentf}
    G = \frac{\bra{1} Te^{-i\hat{H}t} O_N .. O_1 \ket{\rho_0}}{\bra{1}e^{-i\hat{H}t}\ket{\rho_0}} = \bra{1} O(t_N) .. O(t_1) \ket{\rho_0},
\end{equation}
where the imaginary axis propagation is represented by the non-interacting thermal state $\ket{\rho_0}$ and the contour runs only along the real time axis from $t_i$ to $t_f$. The second equality in Eq.~(\ref{greentf}) follows from the trace-preservation of the super-Hamiltonian, allowing us to express the thermal average in an explicitly connected form. In our density-matrix coupled-cluster method, we rigorously impose the trace-preservation of the time-evolved density matrix, resulting in explicit cancellation of the disconnected contributions to the thermal average, which avoids the construction of a correlated bra-state. 

Other choices for the path contour in the complex time plane are also possible because of the analytic property of thermal averages. These choices lead to equivalent formulations of thermal averages in thermo-field theory but may not lead to equivalent coupled-cluster approximations. One other choice, for instance, is to place the backward portion of the real-time propagation at imaginary time $-i\beta/2$ rather than immediately below the real time axis.\cite{matsumoto1984equivalence} This contour gives unitary formulation of thermal averages that has bra-states Hermitian conjugate to the ket-states. If we express the time-dependent density matrix in this unitary formulation in an exponential form, then we also need to compute the correlated bra-state to evaluate the thermal average, unless we use a unitary form for the cluster operator.

We derive the equation of motion for the cluster amplitudes by using Eq.~(\ref{dmcc}) in the Schr\"odinger equation Eq.~(\ref{se}) for the density matrix ket-state.\cite{crawford2000introduction} Following multiplication from the left with $e^{-T_t}$, we obtain
\begin{equation}\label{teom}
    ie^{-T_t}\partial_te^{T_t} \ket{\rho_0} = e^{-T_t}\hat{H}e^{T_t} \ket{\rho_0},
\end{equation}
where $\hat{H}$ is the super-Hamiltonian of the system expressed in the thermal quasi-particle representation. Because $\partial_t$ is a single-particle operator and the creation quasi-particle operators are time-independent, the Baker-Campbell-Hausdorff identity gives
\begin{equation}
    e^{-T_t}\partial_te^{T_t} = \partial_t + \Dot{T}_t,
\end{equation}
where the dot stands for the time derivative. Because the super-Hamiltonian has a finite number of creation and annihilation quasi-particle operators and the cluster operators, Eqs.~(\ref{tser}) and ~(\ref{tbop}), commute, application of the Baker-Campbell-Hausdorff identity shows that the expansion of the similarity-transformed super-Hamiltonian $e^{-T_t}\hat{H}e^{T_t}$ contains a finite number of nested commutator terms. For the standard electronic structure super-Hamiltonian, which has terms with at most four annihilation quasi-particle operators, the commutator series exactly truncates at fourth order. Thus, the equation of motion for the cluster operators, Eq.~(\ref{teom}), simplifies to
\begin{equation}\label{teom2}
    i \Dot{T}_t \ket{\rho_0} = \left( \hat{H} \sum^{n=N}_{n=0} \frac{1}{n!} T^n_t \right)_c \ket{\rho_0},
\end{equation}
where $N$ stands for the largest order non-vanishing term in the series, and we used the time-independence of the ket-state $\partial_t \ket{\rho_0} = 0$. The subscript "c" in Eq.~(\ref{teom2}) denotes that only the connected part of the expression in the brackets has to be considered as a result of the cancellation of the disconnected part by the nested commutators. To derive differential equations for the cluster amplitudes, we project Eq.~(\ref{teom2}) onto states with finite number of quasi-particle excitations, $\bra{0} \tilde{b}_j .. b_i$,
\begin{equation}
    i \bra{0} \tilde{b}_j .. b_i \Dot{T}_t \ket{\rho_0} = \bra{0} \tilde{b}_j .. b_i \left( \hat{H} \sum^{n=N}_{n=0} \frac{1}{n!} T^n_t \right)_c \ket{\rho_0}.
\end{equation}
In our singles and doubles approximation to the DMCC method, Eq.~(\ref{dmccsd}), we project onto bra-states that have only single and double quasi-particle excitations. These equations contain vacuum expectation values of strings of creation and annihilation operators, which when evaluated give a set of coupled non-linear ordinary differential equations for the cluster amplitudes that we solve using the Runge-Kutta method.\cite{press2007numerical}

A distinct advantage of the thermal quasi-particle representation is the existence of traditional normal ordering of operators and the applicability of Wick's theorem for evaluation of expectation values of strings of creation and annihilation operators.\cite{blaizot1986quantum} We define normal-ordering as the string of operators, which has all creation operators to the left of all annihilation operators. Contractions of quasi-particle operators, the difference of any operator ordering from the normal order, are straightforward to evaluate because the quasi-particle operators satisfy the usual commutation relations. Given that the creation operators $b^{\dagger}_i$ and $\tilde{b}^{\dagger}_i$ annihilate the left vacuum $\bra{1}$, vacuum expectation values of normal-ordered strings of operators are identically zero.  Using Wick's theorem, we can thus show that the vacuum expectation value of any string of creation and annihilation quasi-particle operators is equal to the sum of all possible pairwise contractions. If sub-strings are already in normal-ordered form, then contractions need to be considered only among the normal-ordered sub-strings. The number of contractions, however, grows quickly with the complexity of the operator products and diagrammatic representations, which express the contributions to the coupled-cluster equations in a graphical form, become useful. The same set of rules for drawing zero-temperature time-independent coupled-cluster diagrams\cite{bartlett2007coupled,crawford2000introduction} applies to the density-matrix coupled-cluster method with upward going lines representing non-tilde particles and downward going lines -- tilde particles according to the interpretation of non-tilde particles as thermal quasi-particles and tilde particles as thermal quasi-holes. Summation over "internal" lines runs over the entire single-particle basis, and the sign of diagrams is computed by $(-1)^l$, where $l$ is the number of loops determined in the usual way. The direct applicability of the diagrammatic techniques provides powerful methods for evaluating real-time finite-temperature coupled-cluster approximations.

\subsection{Single-impurity Anderson model}\label{siam}
We apply the DMCC method with single and double quasi-particle excitations to the single-impurity Anderson model (SIAM), which is a well-established system to understand the role of electron correlation in dynamical processes at surfaces. SIAM consists of a localized impurity with a repulsive interaction between opposite spin electrons that is coupled to a non-interacting electronic bath.\cite{newns1969self,anderson1961localized,wilson1975renormalization} The SIAM Hamiltonian $H=H_0+W$ has a non-interacting part $H_0$ given by
\begin{equation}\label{siam0}
    H_0 = \sum^{N_b}_{i=0s} \epsilon_{i} a^{\dagger}_{is} a_{is} + \sum^{N_b}_{i=1,s} \left( V_i a^{\dagger}_{0s} a_{is} + a^{\dagger}_{is} a_{0s} \right),
\end{equation}
where $\epsilon_i$ are the single-particle energy levels, $V_i$ -- the impurity-bath coupling elements, $i=0,...,N_b$ enumerates the single-particle basis where $i=0$ corresponds to the impurity and $N_b$ is the number of bath levels, and $s$ sums over the spins of the electrons. The interaction $W$ is given by
\begin{equation}\label{siami}
    W = U a^{\dagger}_{0\alpha}a_{0\alpha}a^{\dagger}_{0\beta}a_{0\beta}
\end{equation}
with $U$ the Hubbard parameter, a measure of the interaction strength. Eqs.~(\ref{siam0}) and ~(\ref{siami}) represent a closed system with a super-Hamiltonian that is the difference of $H$ and its tilde conjugate, $\hat{H} = H - \tilde{H} = H_0 - \tilde{H}_0 + W - \tilde{W}$. Dissipative coupling of the bath levels to a continuum can be incorporated in the SIAM super-Hamiltonian via terms that mix non-tilde and tilde operators,\cite{arimitsu1985general} such as
\begin{equation}\label{diss}
\begin{split}
    \hat{D} = -i \sum^{N_b}_{i=1,s} & [ (\gamma^{(1)}_{i} - \gamma^{(2)}_{i}) \left( a^{\dagger}_{is} a_{is} + \tilde{a}^{\dagger}_{is} \tilde{a}_{is} \right) \\ 
     & - 2 \gamma^{(1)}_{i} \tilde{a}_{is} a_{is} + 2 \gamma^{(2)}_{i} \tilde{a}^{\dagger}_{is} a^{\dagger}_{is} + 2 \gamma^{(2)}_{i} ],
\end{split}
\end{equation}
with $\gamma^{(1)}_{i} = \gamma (1-v_i)$, $\gamma^{(2)}_{i} = \gamma v_i$, and $\gamma$ the dissipation strength. The relation among $\gamma^{(1)}_{i}$, $\gamma^{(2)}_{i}$ and $\gamma$ follows from the detailed balance condition, and $\hat{D}$ conforms with the dissipative part of a Lindblad operator\cite{lindblad1976generators} that represents open systems with Markovian dynamics. Non-Markovian effects of the impurity dynamics are captured by explicitly incorporating a finite electronic bath. Together with dissipation, we include driving in SIAM via a time-dependent modulation of the electronic level of the impurity, $H_t = \epsilon_0 + \delta\epsilon \text{sin}(\Omega t)$. We use logarithmic discretization of the electronic bath in our calculations as suggested by Wilson for the Kondo model\cite{wilson1975renormalization} with a discretization parameter $\Lambda=1.1$ and $N_b = 100$. We consider also the more challenging asymmetric case of SIAM with $\epsilon_0 = -0.08$ eV, impurity-bath coupling strength $V=0.04$ eV, and a range of interaction strengths $U$ and temperatures $T$.

Using the thermal Bogoliubov transformation, Eq.~(\ref{btrans}), the SIAM super-Hamiltonian $\hat{H} = \hat{H}_0 + \hat{W} + \hat{D}$ becomes $\hat{H^\prime} = \hat{H^\prime_0} + \hat{W^\prime}$, where the non-interacting super-Hamiltonian $\hat{H^\prime_0}$ is given by
\begin{equation}\label{hp0}
    \hat{H^\prime_0} = \sum_{i,j,s} h_{ij} b^{\dagger}_{is} b_{js} - h^*_{ijs} \tilde{b}^{\dagger}_{is} \tilde{b}_{js} + \Delta_{ij} \left( b^{\dagger}_{is} \tilde{b}^{\dagger}_{js} - b^{\dagger}_{js} \tilde{b}^{\dagger}_{is} \right),
\end{equation}
with the single-particle Hamiltonian matrix $h_{ijs}$,
\begin{equation}\label{hcore}
\begin{split}
    h_{ij} = \left( \epsilon + v_0 U \right) & \delta_{ij=0} + \epsilon^\prime_i \delta_{ij>0} \\ & + V_i \delta_{i>0, j=0} + V_j \delta_{i=0,j>0},
\end{split}
\end{equation}
and the pairing matrix $\Delta_{ijs}$,
\begin{equation}
    \Delta_{ij} = V_i (u_i v_0 - u_0 v_i) \delta_{i>0, j=0} + V_j (u_j v_0 - u_0 v_j) \delta_{i=0, j>0},
\end{equation}
and the interaction $\hat{W^\prime}$ is written as
\begin{equation}\label{uint}
\begin{split}
    \hat{W^\prime} = & U (u_0-v_0) b^{\dagger}_\alpha b_\alpha b^{\dagger}_\beta b_\beta + U b^{\dagger}_\alpha b_\alpha \tilde{b}_\beta b_\beta + \\ & U \tilde{b}_\alpha b_\alpha b^{\dagger}_\beta b_\beta + U u_0 v_0 b^{\dagger}_\alpha b_\alpha b^{\dagger}_\beta \tilde{b}^{\dagger}_\beta + \\ & U u_0 v_0 b^{\dagger}_\alpha \tilde{b}^{\dagger}_\alpha b^{\dagger}_\beta b_\beta - \text{t.c.}  
\end{split}
\end{equation}
with $\text{t.c.}$ standing for tilde conjugation. In Eq.~(\ref{hcore}), $\epsilon^\prime_i = \epsilon_i - i \gamma$ are the broadened levels of the electronic bath due to dissipation, and the single-particle energy of the impurity is modified by a thermal mean-field term $v_0U$. The Hubbard interaction in Eq.~(\ref{uint}) also incorporates thermal effects as the interaction strength depends on the thermal population of the impurity. Because $\hat{W^\prime}$ includes contributions with up to three annihilation operators, the series in the equation of motion for the cluster operator, Eq.~(\ref{teom2}), truncates at third order, simplifying the differential equations for the cluster amplitudes. In our quantum quench simulations, the impurity is initially decoupled from the bath, $V=0$. Thus, the thermal quasi-particles are defined with respect to the non-interacting thermal state of the decoupled system. At time $t=0$, the coupling to the electronic bath is turned on, and we compute the non-equilibrium electronic dynamics of the impurity.  

\subsection{Finite-temperature renormalization-group approach for open systems}
The quasi-particle representation of thermo-field theory is a suitable framework to extend time-dependent renormalization-group methods\cite{vidal2003efficient,vidal2004efficient,white2004real,schollwock2011density} to closed and open systems at finite temperature. We apply the time-evolving block-decimation approach\cite{vidal2003efficient,vidal2004efficient} to SIAM and compare the results to DMCC in Sec.~\ref{results}. We represent the time-evolving density-matrix ket-state as a matrix product state, for which we arrange the electronic sites in a linear array, starting with the $\alpha$ impurity site, followed by $N_b$ $\alpha$ bath sites, the $\beta$ impurity site, and $N_b$ $\beta$ bath sites. Each electronic site carries both non-tilde and tilde degrees of freedom. The terms of the SIAM super-Hamiltonian $\hat{H^\prime}$ act on two sites at a time with the interaction $\hat{W^\prime}$, Eq.~(\ref{uint}), coupling the impurity sites and the non-interacting super-Hamiltonian $\hat{H^\prime_0}$, Eq.~(\ref{hp0}), acting between the impurity site and each of the bath sites. The super-Hamiltonian terms, however, are non-local as they couple distant sites of the linear array, and we use fermionic swap operators $\hat{S}$ to rearrange sites given by
\begin{equation}
    \hat{S}_{i+1,i} = s_{i+1,i,\alpha} s_{i+1,i,\beta} - \tilde{s}_{i+1,i,\alpha} \tilde{s}_{i+1,i,\beta},
\end{equation}
where
\begin{equation}
    s_{pq\sigma} = 1 + a^{\dagger}_{p,\sigma} a_{q,\sigma} + a^{\dagger}_{q,\sigma} a_{p,\sigma} - a^{\dagger}_{p,\sigma} a_{p,\sigma} - a^{\dagger}_{q,\sigma} a_{q,\sigma}.
\end{equation}
We apply the second-order Suzuki-Trotter decomposition\cite{suzuki1985decomposition} to the super-Hamiltonian terms in the time-evolution operator, which together with the swap operators results in the following sequence of two-site gates for a half time-step $\delta t$
\begin{equation}\label{trotter}
\begin{split}
    e^{-i\hat{H^\prime}\frac{\delta t}{2}} = \prod_{i \in \alpha} \left[e^{-i(\hat{H^\prime_0})_{0i}\frac{\delta t}{2}}\hat{S}_{i,i+1}\right] & e^{-i\hat{W^\prime}\frac{\delta t}{2}} \\
    & \prod_{j \in \beta} \left[e^{-i(\hat{H^\prime_0})_{0j}\frac{\delta t}{2}}\hat{S}_{j,j+1}\right],
\end{split}
\end{equation}
where $i$ runs over the sequence of $\alpha$ sites and $j$ over the sequence of $\beta$ sites, respectively.
The full evolution operator is the product of Eq.~(\ref{trotter}) and its transpose. The application of each of the two-site gates in Eq.~(\ref{trotter}) requires the singular value decomposition of the resultant tensor to obtain the time-evolved matrix product state. Only singular values above a certain threshold are kept in the matrix product state, which represents the decimation step of the renormalization-group algorithm. At zero temperature, the doubled set of tilde and non-tilde degrees of freedom is not necessary because we can evolve only the wavefunction of the system. We apply the same Suzuki-Trotter decomposition using the SIAM Hamiltonian, instead of the super-Hamiltonian, in this case.

\section{Results}\label{results}
\begin{figure}
\includegraphics[width=0.85\linewidth]{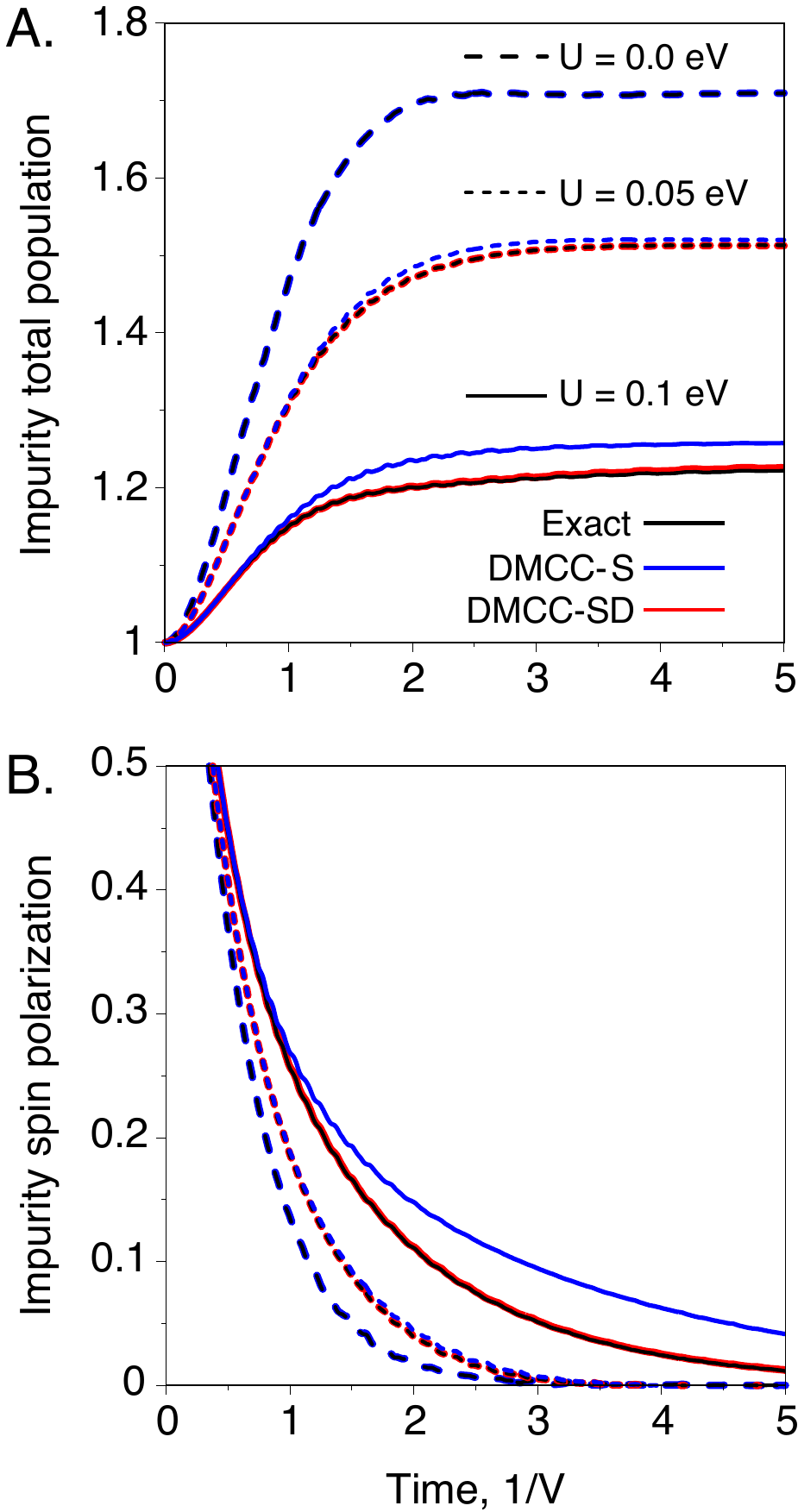}
\caption{\label{fig:fig1} Non-equilibrium dynamics of the single-impurity Anderson model at $T=0$ eV and three values of the Hubbard parameter. Long dashed lines depict the non-interacting $U=0.0$ eV case, short dashed lines  -- the $U=0.05$ eV case, and full lines -- the $U=0.1$ eV case. The renormalization group (Exact) results are plotted in black, the density-matrix coupled-cluster singles (DMCC-S) results are in blue, and the density-matrix coupled-cluster singles and doubles (DMCC-SD) results are in red. (A) Time-dependent total impurity population. (B) Time-dependent impurity spin polarization. Other parameters of the impurity model include $\epsilon_0 = -0.08$ eV, $V = 0.04$ eV, $N_b = 100$, and $\Lambda = 1.1$. For the non-interacting case, the DMCC-S results coincide with the exact results, and for both interacting cases, DMCC-SD is graphically indistinguishable from the exact results.}
\end{figure}

\begin{figure}
\includegraphics[width=0.85\linewidth]{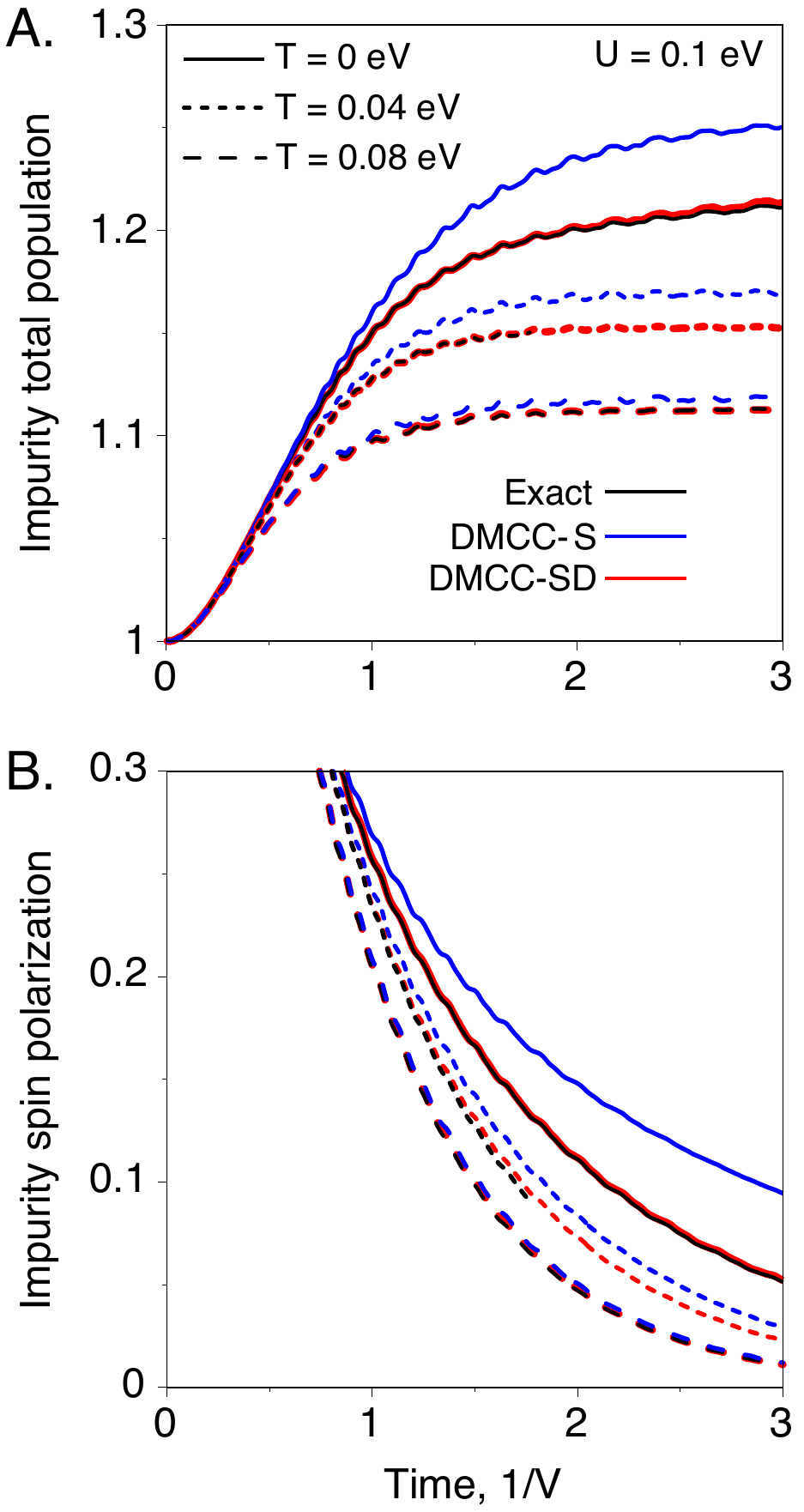}
\caption{\label{fig:fig2} Non-equilibrium dynamics of the single-impurity Anderson model at three values of the temperature and $U=0.1$ eV. Long dashed lines depict high temperature $T=0.08$ eV case, short dashed lines  -- the intermediate $T=0.04$ eV case, and full lines -- the $T=0.0$ eV case. The color code of the curves is the same as in Fig.~\ref{fig:fig1}, as well as the values of the rest of SIAM parameters. (A) Time-dependent total impurity population. (B) Time-dependent impurity spin polarization. The DMCC-S, DMCC-SD and exact results coincide in the high temperature case $T=0.08$ eV.}
\end{figure}

We present the DMCC results for the electronic dynamics of SIAM and compare them with the renormalization-group approach, which we regard as numerically exact. We carry out quantum quench simulations where the impurity is initially decoupled from the bath and is occupied by a single electron, a state with unit total population and unit spin polarization. We turn the coupling to the bath on at $t=0$ and monitor the electronic population dynamics of the impurity. 

Figure ~\ref{fig:fig1} shows the time evolution of the total population (A) and spin polarization (B) of the impurity for three values of the interaction strength $U$ at zero temperature $T=0$. The DMCC singles (DMCC-S) approximation is exact for the non-interacting SIAM $U=0$ as demonstrated by the correspondence to the reference data. With increasing interaction strength, the DMCC-S approximation overestimates the total population and spin polarization of the impurity with larger deviations at higher Hubbard parameter values, revealing the importance of correlation effects. Inclusion of double quasi-particle excitations via the DMCC singles and doubles (DMCC-SD) approximation improves the electronic population dynamics of the impurity. Both the total population and spin polarization of the impurity decrease as a result of correlated dynamics, which brings the coupled-cluster performance in close correspondence to the exact renormalization-group results.

Figure \ref{fig:fig2} depicts the time evolution of the total population (A) and spin polarization (B) of the interacting impurity ($U=0.1$ eV) at three values of temperature $T$. Whereas the zero-temperature SIAM can also be simulated via the time-dependent coupled-cluster wavefunction method, the finite temperature cases require a density-matrix treatment. Both the total population and spin polarization of the impurity decrease with increasing temperature. Similarly, the correlation contribution to the electronic dynamics of the impurity decrease with temperature as measured by the DMCC-S deviation from the reference results. Thus, the DMCC-S approximation, which is equivalent to the time-dependent finite-temperature mean-field method, accurately describes the population dynamics of the impurity at high temperatures. Correlation effects, however, are significant in the low and intermediate temperature regimes and the close correspondence of DMCC-SD with the exact results demonstrates that the method successfully captures the correlation contribution at finite temperatures, as well.

\begin{figure}
\includegraphics[width=0.85\linewidth]{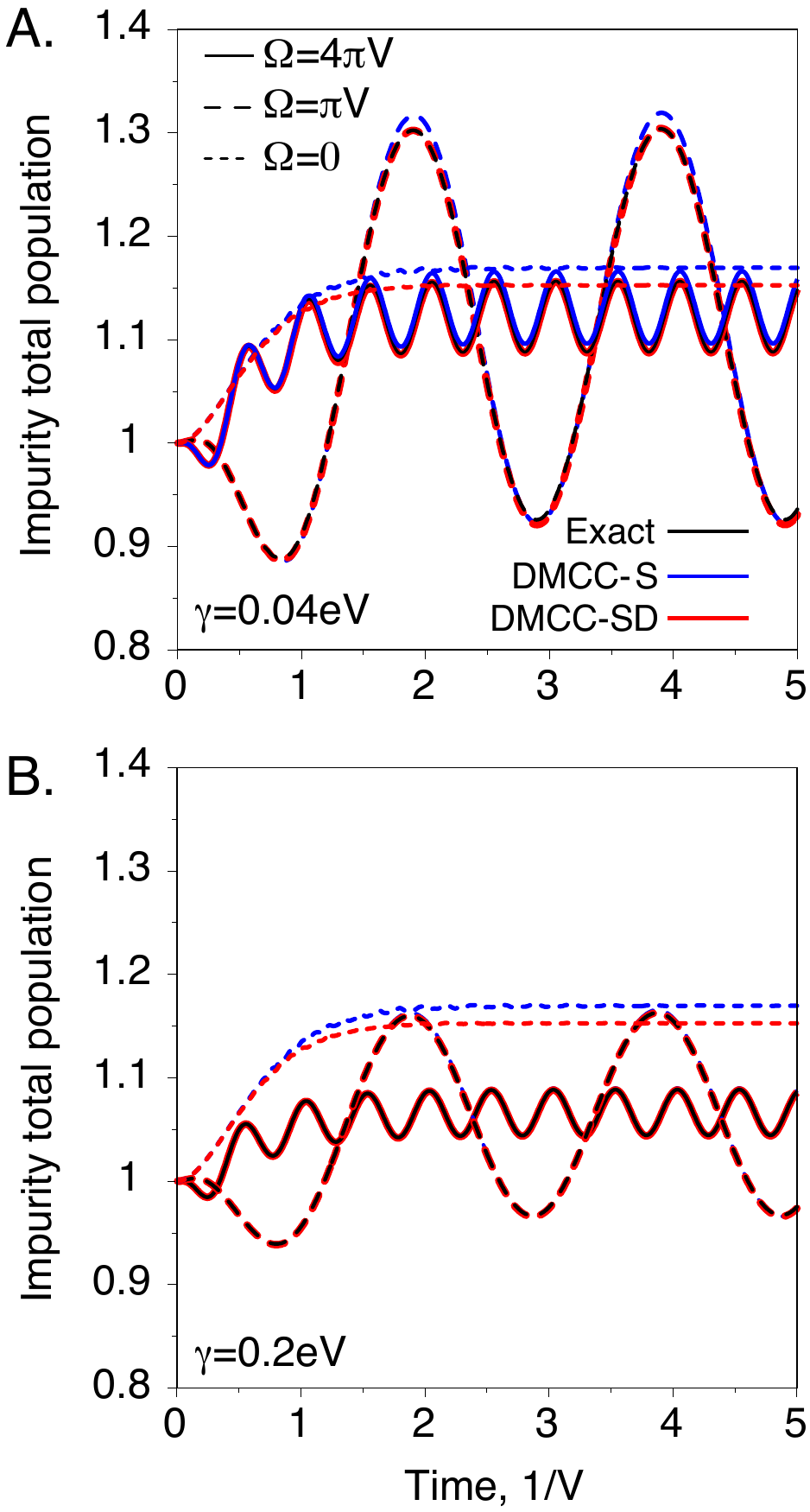}
\caption{\label{fig:fig3} Non-equilibrium dynamics of the single-impurity Anderson model at $T=0.04$ eV and $U=0.1$ eV, including driving and dissipation. We consider two values of the frequency of impurity level modulation, $\Omega$, and dissipation strength, $\gamma$. Full lines depict the high frequency $\Omega=4 \pi V$ case, and long dashed lines -- the low frequency $\Omega=\pi V$ case. (A) Time-dependent total impurity population at $\gamma=0.04$ eV. (B) Time-dependent total impurity population at $\gamma=0.2$ eV. The DMCC-S and the DMCC-SD results without driving and dissipation are also shown in short dashed lines for reference. The color code of the curves is the same as in Fig.~\ref{fig:fig1}, as well as the rest of the SIAM parameters. In (B), the DMCC-S results are indistinguishable from both the DMCC-SD and exact data when dissipation is present. }
\end{figure}
\begin{figure}
\includegraphics[width=0.85\linewidth]{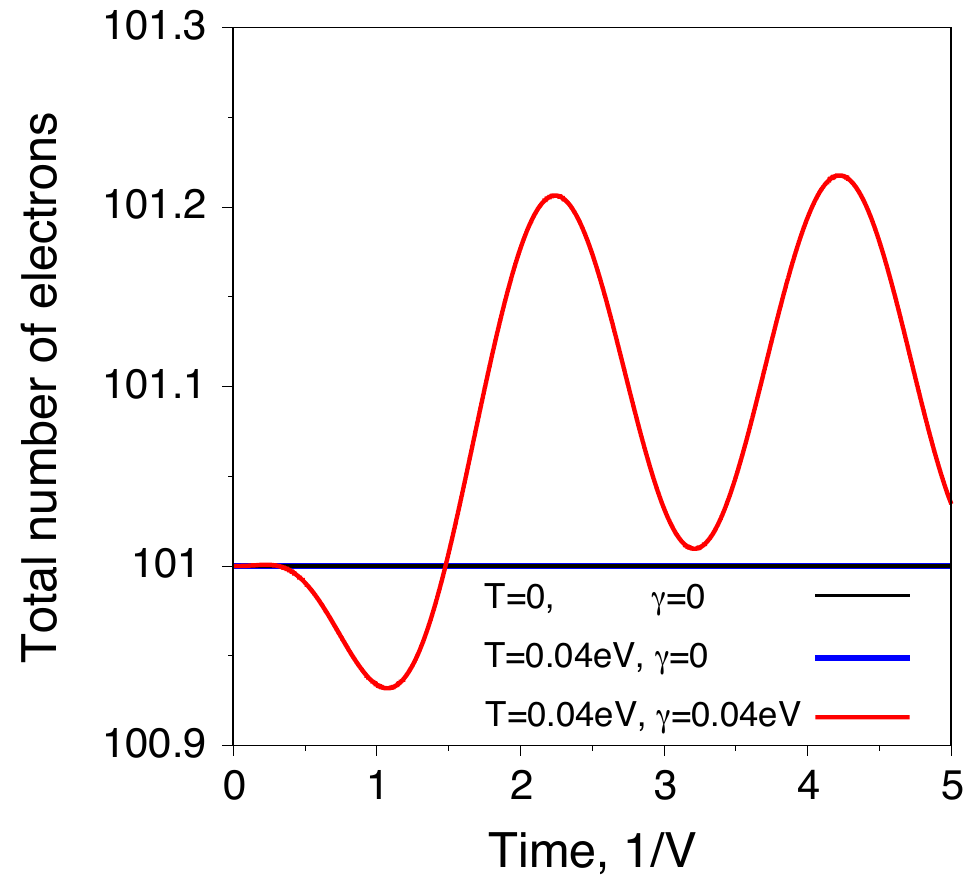}
\caption{\label{fig:fig4} Demonstration of total electron number conservation during the DMCC-SD dynamics of the single-impurity Anderson model. Three cases are considered: zero-temperature closed system ($T=0.0$ eV, $\gamma=0.0$ eV) in black, finite-temperature closed system ($T=0.04$ eV, $\gamma=0.0$ eV) in blue, and, finite-temperature open system ($T=0.04$ eV, $\gamma=0.04$ eV) in red. The closed system results (black and blue lines) are graphically indistinguishable. }
\end{figure}

The models considered so far represent closed quantum systems with dynamics characterized by relaxation to the equilibrium state of the interacting system. The DMCC method also allows us to investigate the electronic dynamics of driven-dissipative open quantum systems. In particular, we carry out quantum quench simulations of the driven-dissipative SIAM with a super-Hamiltonian that includes a dissipative contribution, Eq.~(\ref{diss}), and a harmonic driving contribution, $H_t = \epsilon_0 + \delta \epsilon \text{sin} (\Omega t)$ with $\delta \epsilon = 0.08$ eV. Figure \ref{fig:fig3} shows the total population dynamics of the interacting impurity ($U=0.1$ eV) at temperature $T=0.04$ eV and two values of the modulation frequency, $\Omega=\pi V$ and $\Omega=4 \pi V$. Figure \ref{fig:fig3}A also includes dissipation with strength $\gamma=0.04$ eV, and Figure \ref{fig:fig3}B -- dissipation with strength $\gamma=0.2$ eV.
Figure \ref{fig:fig3}A reveals that the amplitude of impurity population oscillations decreases with increasing modulation frequency. This non-adiabatic effect results from the inefficient equilibration of the impurity with the electronic bath, and is reproduced by both DMCC approximations. The DMCC-S results, however, show asymmetric deviations from the reference data, especially in the low-frequency case where the DMCC-S deviations are larger near the peaks of the population oscillations. This time-evolving correlation effect is captured by DMCC-SD and results from the decrease of the correlation contribution as the impurity depopulates. Comparison of the results in Figures \ref{fig:fig3}A and \ref{fig:fig3}B further shows that both the amplitude of oscillations and average value of impurity population decrease with increasing dissipation strength. The coincidence of DMCC-S, DMCC-SD and exact results demonstrates that dissipation depresses correlation effects, similarly to increasing temperature. Overall, the DMCC-SD approximation reproduces well the rich non-equilibrium behavior of the model, underscoring the promise of the approach to study systems far from equilibrium. 

Finally, Figure \ref{fig:fig4} demonstrates the particle number conservation during the coupled-cluster time evolution. The number of electrons is conserved within numerical convergence thresholds for the closed system at zero and finite temperature, and shows characteristic oscillations for the driven-dissipative, open quantum system. Taken together, the DMCC-SD method successfully captures the non-equilibrium many-body dynamics of SIAM in close correspondence with the renormalization-group method.

\section{Conclusions}\label{conclusion}
The current work presents an extension of coupled-cluster theory to real-time finite-temperature correlated quantum dynamics. The method uses the Keldysh path contour in the complex time plane, which incorporates thermal correlation effects via thermalization from a non-interacting thermal state. The density matrix is expressed in an exponential ansatz and the equation of motion for the cluster amplitudes is obtained via projection from the Schr\"odinger equation for the cluster operator. The thermal quasi-particle presentation of thermo-field theory is applied to the cluster operator, ensuring the trace-preservation of the density matrix. In addition, the diagrammatic techniques of the traditional coupled-cluster approach are directly transferable, which facilitates the computational implementation of the method.

Investigation of the single-impurity Anderson model demonstrates the accuracy of the approach in reproducing the electronic dynamics of the impurity compared to renormalization-group methods. The coupled-cluster time evolution captures the balance of correlation effects with increasing interaction strength and temperature in closed systems, and also efficiently includes driving and dissipation effects in open systems. Thus, the method provides a unified framework to simulate the non-equilibrium many-body dynamics of both closed and open systems at finite temperature.

This work presents a study of a coupled-cluster formalism with many avenues for further developments. One possibility is the more general definition of the thermal Bogoliubov transformation. Furthermore, the time-dependent approach can serve as a basis for more approximate treatments that explicitly include thermal and dissipative effects, and possibly more general single-particle density matrices can also be accommodated. 

\begin{acknowledgments}
P.S. would like to gratefully acknowledge financial support by the German Science Foundation.
\end{acknowledgments}

\nocite{*}
\bibliography{dmcc.bib}

\end{document}